\documentclass[12pt]{article}

\newcommand{\qed}{{\hfill\rule{4pt}{7pt}}}

\usepackage{amsmath, epsfig}
\usepackage{amssymb}
\usepackage{amsfonts}
\usepackage{latexsym}
\usepackage{color}

\newtheorem{thm}{Theorem}[section]
\newtheorem{prop}[thm]{Proposition}

\newtheorem{cor}[thm]{Corollary}

\def\pf{\noindent {\it Proof.} }

\numberwithin{equation}{section} \numberwithin{figure}{section}

\def\qed{\hfill \rule{4pt}{7pt}}

\def\pf{\noindent {\it Proof.} }

\begin{document}

\begin{center}

{\large\bf A Reflection Principle for Three Vicious Walkers  }

\vskip 2mm

William Y. C. Chen$^{1}$,
 Donna Q. J. Dou$^{2}$,   Terence Y. J. Zhang$^{3}$

\vskip 2mm

 Center for Combinatorics, LPMC-TJKLC\\
 Nankai
University\\
Tianjin 300071, P.R. China

 \vskip 0.2 cm $^1$chen@nankai.edu.cn,
$^2$qjdou@cfc.nankai.edu.cn, $^3$terry@mail.nankai.edu.cn

\end{center}

\begin{abstract} We establish a reflection principle for three
lattice walkers and  use this principle to reduce the enumeration of
the configurations of three vicious walkers to that of
configurations of two vicious walkers. In the combinatorial
treatment of two vicious walkers, we make connections to two-chain
watermelons and to the classical ballot problem. Precisely, the
reflection principle leads to a bijection between three walks $(L_1,
L_2, L_3)$ such that $L_2$ intersects  both $L_1$ and $L_3$ and
three walks $(L_1, L_2, L_3)$ such that $L_1$ intersects  $L_3$.
Hence we find a combinatorial interpretation of the formula for the
generating function for the number of configurations of three
vicious walkers, originally derived by Bousquet-M\'elou by using the
kernel method, and independently by Gessel  by using tableaux and
symmetric functions.

\end{abstract}

\noindent {\bf Keywords:} vicious walkers, watermelon, Catalan
numbers, Ballot numbers, reflection principle.

\noindent {\bf AMS Classification Numbers:} 82B23; 05A15

\section{Introduction}

The vicious walker model was  introduced by Fisher \cite{Fish} in
1984 and has drawn much attention. A walker is said to be vicious if
he does not like to meet any other walker at any point.  Formally
speaking, a configuration of $r$ vicious walkers, called $r$ vicious
walks, of length  $n$, is an $r$-tuple of pairwise nonintersecting
lattice walks of length $n$, consisting of up steps $U$ (i.e.,
$(1,1)$) and down steps $D$ (i.e., $(1,-1)$), starting from
$(0,2i_1), (0,2i_2), \ldots, (0,2i_r)$ and ending at
$(n,e_1),(n,e_2),\ldots, (n,e_r)$ where $i_r>\cdots> i_2>i_1=0$ and
$e_r>\cdots >e_2>e_1$. Precisely, two lattice paths are said to be
nonintersecting if they do not share any common points. In
particular, a watermelon of length $n$ is a configuration consisting
of $r$ chains, or paths, of length $n$ which start at the points
$(0,0), (0,2), \ldots, (0,2r-2)$ and end at the points
$(n,k),(n,k+2),\ldots, (n,k+2r-2)$ for some $k$. In other words, a
watermelon is a vicious walker configuration starting at adjacent
points and ending at adjacent points. Note that two lattice points
are said to be adjacent if they are on the same vertical line and
their $y$-coordinates  differ by $2$. It is known that
configurations of vicious walkers can be represented by tableaux. So
the theory of symmetric functions can be employed to study vicious
walkers, see \cite{Ges-Vie, Gutt1,Gutt4, Hika,Gutt2,Gutt3}.

The main objective of this paper is to present a combinatorial
approach to the enumeration of configurations of three vicious
walkers. Let us fix the starting points $(0,0)$, $(0, 2i)$ and $(0,
2i+2j)$. Let $V(i,j,n)$ be the set of three vicious walks $(L_1,
L_2, L_3)$ of length $n$, where $L_1$ is the path of the first
walker starting from $(0,0)$, $L_2$ is the path of the second walker
starting from $(0, 2i)$, and $L_3$ is the path of the third walker
starting from $(0, 2i+2j)$. Define the generating function
$V_{i,j}(t)$ to be
\begin{equation} \label{dvijt}
V_{i,j}(t) =\sum_{n=0}^\infty \, |V(i,j,n)| t^n,
\end{equation}
where $|\cdot|$ denotes the cardinality of a set.

The  enumeration of configurations of three vicious walkers has been
solved independently by Bousquet-M\'elou \cite{Melou} by using the
kernel method, and by Gessel \cite{Gess} by using tableaux and
symmetric functions. They obtained a formula for $V_{i,j}(t)$ in
terms of the generating function of the Catalan numbers.

Let $C(t)$ be the generating function of the Catalan numbers
$C_n={1\over n+1} {2n\choose n}$, that is,
\[ C(t)=\sum_{n=0}^\infty C_n t^n.\]
Recall that $C(t)$ satisfies the recurrence relation
\begin{equation}
\label{cT-eq}
    C(t) = 1 +t C^2(t).
\end{equation}
Let
 \begin{equation} \label{dtc}
 D(t)=tC^2(t)=C(t)-1=\sum_{n=0}
C_{n+1}t^{n+1}.
\end{equation}

The following elegant formula is due to Bousquet-M\'{e}lou
\cite{Melou} and Gessel \cite{Gess}.

\begin{thm}[Bousquet-M\'elou \cite{Melou} and Gessel
\cite{Gess}]\label{mainthm2}
\begin{equation}\label{maineqn2}
V_{i,j}(t)={1\over 1-8t} (1-D^i(2t))(1-D^j(2t)).
\end{equation}
\end{thm}

In view of the relation (\ref{dtc}) and the identity
\begin{equation}\label{d2t}
\Big(\frac{1+D(t)}{1-D(t)}\Big)^2= { 1\over 1-4t},
\end{equation}
Gessel derived the following form of the formula for $V_{i,j}(t)$.

\begin{thm}
[Gessel \cite{Gess}] \label{mainthm} For any $i,j\geq 1$, we have
\begin{equation}\label{maineqn}V_{i,j}(t)
=C^2(2t)\big(1+D(2t)+\cdots+D^{i-1}(2t)\big)
\big(1+D(2t)+\cdots+D^{j-1}(2t)\big).
\end{equation}
\end{thm}

Both  Bousquet-M\'elou \cite{Melou} and Gessel \cite{Gess} proposed
the problem of finding a combinatorial interpretation of the formula
for  $V_{i,j}(t)$. The question of Bousquet-M\'elou is concerned
with the formula (\ref{maineqn2}), while the question of Gessel is
concerned with the formula in the form of (\ref{maineqn}). In this
paper, we will present a combinatorial interpretation of
(\ref{maineqn2}). As will be seen, the algebraic manipulations to
transform the formula (\ref{maineqn2}) to  (\ref{maineqn}) can be
explained combinatorially. So we have obtained combinatorial
interpretations of both formulas (\ref{maineqn2}) and
(\ref{maineqn}).

We also take a different approach to the enumeration of
configurations of two vicious walkers. By reformulating the problem
in terms of pairs of intersecting walks, we give a decomposition of
a pair of converging walks, that is, two walks that do not intersect
until they reach the same ending point,  into two-chain watermelons,
or $2$-watermelons. Then we can use Labelle's formula for the number
of $2$-watermelons of length $n$ to derive the formula for the
number of two vicious walks of length $n$. In the last section, we
make a connection between  pairs of converging walks and the
classical ballot numbers, by applying the Labelle merging algorithm,
in the form presented by Chen, Pang, Qu and Stanley \cite{CPQS},

\section{The Reflection Principle}

In this section, we will establish  a reflection principle so that
we can reduce the enumeration of three vicious walkers to that of
two vicious walkers. This reduction leads to a combinatorial
interpretation of the formula for $V_{i,j}(t)$, as defined by
(\ref{dvijt}).

Let us recall some basic definitions. Two walks $L_1$ and $L_2$ are
said to be intersecting, denoted $L_1\cap L_2\neq \emptyset$, if
$L_1$ and $L_2$ share a common point.
 Let $U(i,j,n)$ be the set of all $3$-walks
$(L_1,L_2,L_3)$ of length $n$, where $L_1$, $L_2$ and $L_3$ start
from $(0,0)$, $(0,2i)$ and $(0,2i+2j)$ respectively. Let
\[U_{i,j}(t) =\sum_{n=0}^\infty \, |U(i,j,n)| t^n.\]
It is obvious that
\begin{equation}\label{uij}
 U_{i,j}(t)={ 1 \over 1-8t}.
 \end{equation}
  We use $W_{12}(n)$, or
$W_{12}$ for short, to denote the set of $3$-walks $(L_1,L_2,L_3)$
in $U(i,j,n)$ such that $L_1$ and $L_2$ are nonintersecting.
Similarly, we use $W_{23}(n)$, or $W_{23}$ for short, to denote the
set of $3$-walks $(L_1, L_2, L_3)$ in $U(i,j,n)$ such that $L_2$ and
$L_3$ are nonintersecting. Clearly, the set $V(i,j,n)$ of three
vicious walks of length $n$ can be expressed as $W_{12}\cap W_{23}$.
By the principle of inclusion and exclusion, we see that
\begin{equation}\label{v-ijn}
|V(i,j,n)|=|W_{12} \cap W_{23}|=|W_{12}|+|W_{23}|-|W_{12}\cup
W_{23}|.
\end{equation}
In order to compute $|W_{12}\cup W_{23}|$, we let $M_{12, 23}(n)$,
or $M_{12, 23}$ for short,  denote the set of $3$-walks
$(L_1,L_2,L_3)$ in $U(i,j,n)$ such that $L_2$ intersects   both
$L_1$ and $L_3$. Clearly, we have
\begin{equation} \label{w1223}
 |W_{12} \cup W_{23}| = |U(i,j,n)| - |M_{12, 23}|.
 \end{equation}

We are now in a position to establish a reflection principle to deal
with the enumeration of $M_{12, 23}(n)$. Let $M_{13}(n)$, or
$M_{13}$ for short, denote the set of $3$-walks $(L_1,L_2,L_3)$ in
$U(i,j,n)$ such that $L_1$ intersects   $L_3$. Then we have the
following correspondence.

\begin{thm} \label{bij} For $n\geq 1$, there
exists a bijection between $M_{12, 23}(n)$ and $M_{13}(n)$.
\end{thm}

\pf We construct a map $\Phi$ from $M_{12, 23}(n)$ to $M_{13}(n)$ as
follows. Let $(L_1,L_2,L_3)$ be a $3$-walk in $M_{12, 23}(n)$. We
consider the following two cases. If $L_1\cap L_3\neq \emptyset$,
then it is clear that $(L_1,L_2,L_3)\in M_{13}(n)$. In this case, we
define
 $\Phi((L_1,L_2,L_3))=(L_1,L_2,L_3)$.

\begin{figure}[h,t]
\setlength{\unitlength}{1pt}
\begin{center}
\begin{picture}(280,80)

\put(-12,0){$L_1$}\put(-12,25){$L_2$}\put(-12,50){$L_3$}
\qbezier(0,0)(20,-10)(40,20)\qbezier(40,20)(60,40)(80,15)\qbezier(80,15)(90,5)(120,20)
\qbezier(0,25)(30,4)(60,40)\qbezier(60,40)(90,60)(120,30)
\qbezier(0,50)(40,90)(90,40)\qbezier(90,40)(105,30)(120,50)
\put(15,-6){1}\put(90,8){1}
\put(15,13){2}\put(60,35){2}\put(92,45){2}\put(118,30){2}
\put(15,57){3}\put(92,35){3}\put(118,48){3}

\put(136,28){$\stackrel{\Phi}\longrightarrow$}

\put(168,0){$L_1^{\prime}$}\put(168,25){$L_2^{\prime}$}\put(168,50){$L_3^{\prime}$}
\qbezier(180,0)(200,-10)(220,20)\qbezier(220,20)(240,40)(260,15)\qbezier(260,15)(270,5)(300,20)
\qbezier(180,25)(210,4)(240,40)\qbezier(240,40)(270,60)(300,30)
\qbezier(180,50)(220,90)(270,40)\qbezier(270,40)(295,30)(300,50)
\put(225,25){\circle*{2}}\put(45,25){\circle*{2}}
\put(220,26){$P$}\put(41,26){$P$}
\put(195,-6){1}\put(270,8){2}
\put(195,13){2}\put(240,35){1}\put(272,45){1}\put(298,30){1}
\put(195,57){3}\put(272,35){3}\put(298,48){3}

\end{picture}
\end{center}
\caption[subsection]{\label{fig-bijection} The reflection
principle.}
\end{figure}
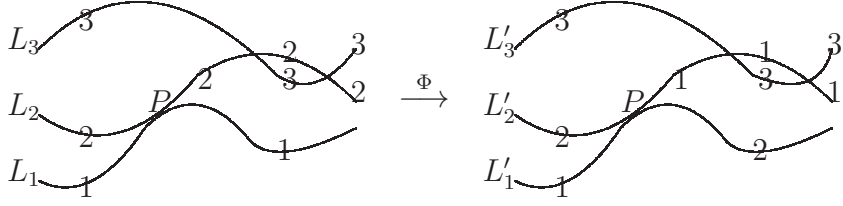

 We may now assume that $L_1\cap L_3=\emptyset$.
 We first consider the
case that $L_2$  meets $L_1$ before it  meets $L_3$. Suppose that
$P$ is the first intersection point of
 $L_2$ and
$L_1$. We now conduct the usual reflection operation on $L_1$ and
$L_2$, and denote the resulting paths by $L_1'$ and $L_2'$. Namely,
$L_1'$ consists of the first segment of $L_1$ up to the point $P$
followed by the last segment of $L_2$ starting from the point $P$,
and $L_2'$ consists of the first segment of $L_2$ up to the point
$P$ followed by the last segment of $L_1$ starting from the point
$P$. Figure \ref{fig-bijection} is an illustration of the
reflection.

Let $L_3'=L_3$ and $\Phi((L_1,L_2,L_3))=(L_1',L_2',L_3')$. It is
clear that $L_1^{\prime}$ must meet $L_3^{\prime}$. Thus we have
$(L_1^{\prime},L_2^{\prime},L_3^{\prime})\in M_{13}(n)$.

It is not difficult to see that the above procedure is
 reversible.
We are still left with the case when $L_2$ intersects   $L_3$ before
meeting $L_1$. This case is analogous to the case that we have
considered. Thus we have reached the conclusion that $\Phi$ is a
bijection.\qed

Combining \eqref{v-ijn}, \eqref{w1223} and Theorem \ref{bij}, we
obtain the following relation
\begin{equation}\label{v-ijn2}
|V(i,j,n)|=|W_{12}|+|W_{23}|+|M_{13}| -|U(i,j,n)|.
\end{equation}

Let $W_{13}$ be the set of three walks $(L_1, L_2, L_3)$ in
$U(i,j,n)$ such that $L_1$ never meets $L_3$, and define the
generating functions for $|W_{12}|$, $|W_{23}|$ and $|W_{13}|$ by
$W_{12}(t)$, $W_{23}(t)$ and $W_{13}(t)$ respectively. From
(\ref{v-ijn2}) it follows that
\begin{equation}\label{reduction 1}
|V(i,j,n)|   =   |W_{12}| + |W_{23}| -|W_{13}|.
\end{equation}

\begin{prop}\label{redution-1}
\begin{equation}\label{reduction-1}
V_{i,j}(t)   =   W_{12}(t) + W_{23}(t) -W_{13}(t).
\end{equation}
\end{prop}

The above formula can be viewed as a reduction of the three vicious
walkers problem to that of two vicious walkers. Let $N(i,n)$ be the
set of two vicious walks $(L_1,L_2)$ of length $n$ starting at
$(0,0)$ and $(0,2i)$ respectively, and denote the corresponding
generating function by
\[N_i(t)=\sum_{n=0}^{\infty}|N(i,n)|t^n.\]

  Bousquet-M\'elou \cite{Melou} and  Gessel  \cite{Gess}
  obtained the
following formula
\begin{equation}\label{tv2}
N_i(t)={1\over 1-4t}(1-D^i(t)).
\end{equation}
As pointed out by Gessel \cite{Gess}, the above formula for
$N_{i}(2t)$ can be deduced from the formula (\ref{maineqn}) for
$V_{i,j}(t)$ by taking the limit $j \rightarrow¡¡\infty$, and by
using the identity (\ref{d2t}).

Using the above formula for $N_i(t)$, one can derive the following
formulas for the generating functions $W_{12}(t)$, $W_{23}(t)$ and
$W_{13}(t)$:
\begin{equation}\label{w123}
W_{12}(t) = {1-D^i(2t)\over 1-8t}, \; W_{23}(t)={1-D^j(2t)\over
1-8t}, \; W_{13}(t)={1-D^{i+j}(2t)\over 1-8t}.
\end{equation}
Clearly, formula \eqref{maineqn2} in Theorem \ref{mainthm2} follows
from the above formulas and the relation  \eqref{reduction-1}.

We note that Gessel \cite{Gess}
 obtained the
following identity
\begin{equation} \label{vn}
V_{i,j}(t)=N_i(2t)+N_j(2t)-N_{i+j}(2t),
\end{equation}
in accordance with the combinatorial statement \eqref{reduction-1}
derived from the reflection principle.

As to the question of finding a combinatorial interpretation of the
generating function formula \eqref{maineqn2},  the reflection
principle (Theorem \ref{bij}) along with the combinatorial
interpretations of the formulas for $W_{12}(t)$, $W_{23}(t)$ and
$W_{13}(t)$ can be considered as an answer because the principle of
inclusion and exclusion for two sets can be easily justified
combinatorially. In the next section, we will present a
combinatorial treatment of the formula \eqref{tv2} for two vicious
walkers. Moreover, we note that one can give a combinatorial
reasoning of the transformation from the formula \eqref{maineqn2} to
the formula \eqref{maineqn}.

It is to deduce \eqref{maineqn} from \eqref{maineqn2} by utilizing
the identity (\ref{d2t}), which can be  explained combinatorially in
two steps. The first step is to show that
\begin{equation}\label{4n}
4^n= \sum_{k=0}^{2n} {2k \choose k}{2n-2k \choose n-k},
\end{equation}
which is equivalent to the identity
\begin{equation}
\sum_{n=0}^\infty {2n \choose n}t^n = {1 \over \sqrt{1-4t}}.
\end{equation}
There are several combinatorial proofs of (\ref{4n}), see, for
example, Kleitman \cite{Klei} and  Marta \cite{SvMa}. The second
step is to show that
\begin{equation} \label{dt} {1+D(t)\over 1-D(t)}
=\sum_{n=0}^\infty {2n \choose n}t^n.
\end{equation}
 Note that ${1+D(t) \over 1-D(t)}$ can be
written as ${C(t) \over 1-tC^2(t)}$.  A  combinatorial
interpretation of the identity
\[
{C(t)\over 1-tC^2(t)}=\sum_{n=0}^\infty {2n \choose n}t^n\] is given
by Chen, Li and Shapiro \cite{CLSh}
 in terms of doubly rooted plane
trees and the butterfly decomposition.

\section{Converging Walks and $2$-Watermelons}

In this section, we present a different approach to the two vicious
walkers problem by counting pairs of converging walks. A pair of
 walks is said to be converging if they never meet until
they reach a common ending point. We will show that pairs of
converging walks can be enumerated by applying Labelle's formula for
two-chain watermelons, or $2$-watermelons \cite{Labe}. Precisely, we
will give a decomposition of a pair of converging walks into
$2$-watermelons.

Recall that $M_{13}(n)$ is defined in the previous section. Let
$M_{12}(n)$, or $M_{12}$ for short, be the set of 3-walks $(L_1,
L_2, L_3)$ in $U(i,j,n)$ such that $L_1$ intersects $L_2$.
Similarly, we can define $M_{23}(n)$, or $M_{23}$ for short.
Clearly, we have
\[ |M_{12}|= |U(i,j,n)|-|W_{12}|, \quad
|M_{23}|=|U(i,j,n)|-|W_{23}|.\] From (\ref{v-ijn2}) it follows that
\[ |V(i,j,n)|=|U(i,j,n)|
+|M_{13}|-|M_{12}|-|M_{23}|. \]

Let $M_{12}(t)$, $M_{23}(t)$ and $M_{13}(t)$ denote the generating
functions for $|M_{12}(n)|$, $|M_{23}(n)|$ and $|M_{13}(n)|$,
respectively.

\begin{prop}We have
\label{intersect to vicious}
\begin{equation} \label{vm}
V_{i,j}(t) = U_{i,j}(t)+M_{13}(t)- M_{12}(t) -M_{23}(t).
\end{equation}
\end{prop}

We will show that $M_{12}(t)$, $M_{13}(t)$ and $M_{23}(t)$ can be
computed by using Labelle's formula for $2$-watermelons.

\begin{prop}[Labelle \cite{Labe}] \label{watermelon}
The number of $2$-watermelons with each walk having $n$ steps is
$C_{n+1}$.
\end{prop}

By Labelle's formula, one sees that the generating function of the
number of $2$-watermelons equals $C^2(t)$. Note that $2$-watermelons
of length $n$ correspond to pairs of converging walks of length
$n+1$ with adjacent starting points.  In general, let $T(i,n)$ be
the set of pairs of converging walks $(L_1,L_2)$ of length $n$,
where $L_1$ starts from $(0,0)$ and $L_2$ starts from $(0, 2i)$.
Define
\begin{equation*} T_i(t)=\sum_{n\geq
0}|T(i,n)|t^n.
\end{equation*}

\begin{prop}\label{b-prop}
For any $i\geq 1$,    $T_i(t)=D^i(t)$.
\end{prop}

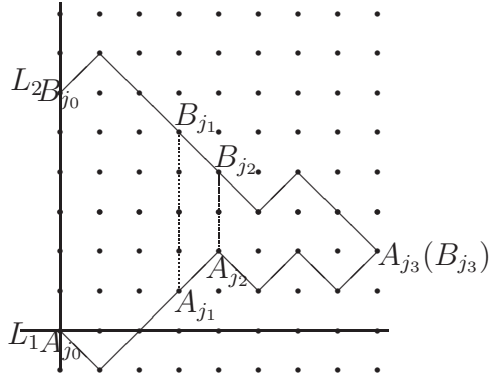
\begin{figure}[h,t]
\setlength{\unitlength}{1pt}
\begin{center}
\begin{picture}(90,130)
\put(-35,15){\line(1,0){139}} \put(-20,0){\line(0,1){139}}

\multiput(-20,0)(15,0){9}{\circle*{2}}
\multiput(-20,15)(15,0){9}{\circle*{2}}
\multiput(-20,30)(15,0){9}{\circle*{2}}
\multiput(-20,60)(15,0){9}{\circle*{2}}
\multiput(-20,45)(15,0){9}{\circle*{2}}
\multiput(-20,60)(15,0){9}{\circle*{2}}
\multiput(-20,75)(15,0){9}{\circle*{2}}
\multiput(-20,90)(15,0){9}{\circle*{2}}
\multiput(-20,105)(15,0){9}{\circle*{2}}
\multiput(-20,120)(15,0){9}{\circle*{2}}
\multiput(-20,135)(15,0){9}{\circle*{2}}
\put(-20,15){\line(1,-1){15}} \put(-5,0){\line(1,1){45}}
\put(40,45){\line(1,-1){15}}\put(55,30){\line(1,1){15}}
\put(70,45){\line(1,-1){15}}\put(85,30){\line(1,1){15}}
\put(-20,105){\line(1,1){15}}\put(-5,120){\line(1,-1){60}}
\put(55,60){\line(1,1){15}}\put(70,75){\line(1,-1){30}}

\put(-40,10){$L_1$}\put(-39,106){$L_2$}
\put(-29,102){$\small{B_{j_0}}$}\put(-28,7){$\small{A_{j_0}}$}
\put(23,93){$\small{B_{j_1}}$}\put(23,22){$\small{A_{j_1}}$}
\put(38,77){$\small{B_{j_2}}$}\put(35,35){$\small{A_{j_2}}$}
\put(100,40){$\small{A_{j_3}(B_{j_3}})$}

\qbezier[40](25,30)(25,60)(25,90)\qbezier[30](40,45)(40,60)(40,75)

\end{picture}
\caption{The decomposition of a pair of converging
walks.\label{decomposition}}
\end{center}
\end{figure}

\pf Let $L_1=A_0A_1\ldots A_n$ and $L_2=B_0 B_1\ldots B_n$, where a
walk is represented by a sequence of points. For $0\leq k\leq i$,
let $j_k$ be the minimum index such that the difference of the
$y$-coordinates of $(A_{j_k},B_{j_k})$ equals to $2i-2k$. It is
clear that $j_0=0$ and $j_i=n$. We now decompose $(L_1,L_2)$  into
$i$ 2-walks: $(L_1^{(1)},L_2^{(1)}),\ldots,(L_1^{(i)},L_2^{(i)})$,
where $L_1^{(k)}=A_{j_{k-1}}A_{j_{k-1}+1} \ldots A_{j_k}$ and
$L_2^{(k)}=B_{j_{k-1}} B_{j_{k-1}+1}  \ldots B_{j_{k}}$.  Figure
\ref{decomposition} is an illustration of the decomposition.

\begin{figure}[h,t]
\setlength{\unitlength}{1pt}
\begin{center}
\begin{picture}(240,110)
\put(-48,12){\line(1,0){48}} \put(-24,0){\line(0,1){96}}

\multiput(-24,0)(12,0){4}{\circle*{2}}
\multiput(-24,12)(12,0){4}{\circle*{2}}
\multiput(-24,24)(12,0){4}{\circle*{2}}
\multiput(-24,36)(12,0){4}{\circle*{2}}
\multiput(-24,48)(12,0){4}{\circle*{2}}
\multiput(-24,60)(12,0){4}{\circle*{2}}
\multiput(-24,72)(12,0){4}{\circle*{2}}
\multiput(-24,84)(12,0){4}{\circle*{2}}
\multiput(-24,96)(12,0){4}{\circle*{2}}

\put(-24,12){\line(1,-1){12}} \put(-12,0){\line(1,1){12}}
\qbezier[8](0,12)(6,18)(12,24)
\put(-24,84){\line(1,1){12}}\put(-12,96){\line(1,-1){12}}
\qbezier[8](0,84)(6,78)(12,72) \put(-42,8){$L_1^{(1)}$}
\put(-42,78){$L_2^{(1)}$}

\put(19,45){$\rightarrow$}

\put(24,12){\line(1,0){36}} \put(36,0){\line(0,1){96}}

\multiput(36,0)(12,0){3}{\circle*{2}}
\multiput(36,12)(12,0){3}{\circle*{2}}
\multiput(36,24)(12,0){3}{\circle*{2}}
\multiput(36,36)(12,0){3}{\circle*{2}}
\multiput(36,48)(12,0){3}{\circle*{2}}
\multiput(36,60)(12,0){3}{\circle*{2}}
\multiput(36,72)(12,0){3}{\circle*{2}}
\multiput(36,84)(12,0){3}{\circle*{2}}
\multiput(36,96)(12,0){3}{\circle*{2}}

\put(36,12){\line(1,-1){12}} \put(48,0){\line(1,1){12}}
\put(36,36){\line(1,1){12}}\put(48,48){\line(1,-1){12}}

\multiput(96,0)(12,0){2}{\circle*{2}}
\multiput(96,12)(12,0){2}{\circle*{2}}
\multiput(96,24)(12,0){2}{\circle*{2}}
\multiput(96,36)(12,0){2}{\circle*{2}}
\multiput(96,48)(12,0){2}{\circle*{2}}
\multiput(96,60)(12,0){2}{\circle*{2}}
\put(84,12){\line(1,0){24}}\put(96,0){\line(0,1){60}}
\qbezier[8](96,12)(102,18)(108,24)
\qbezier[8](96,60)(102,54)(108,48) \put(80,9){$L_1^{(2)}$}
\put(80,57){$L_2^{(2)}$} \put(115,33){$\rightarrow$}
\put(132,33){$\emptyset$}

\put(156,12){\line(1,0){60}}\put(168,0){\line(0,1){48}}
\multiput(168,12)(12,0){5}{\circle*{2}}
\multiput(168,24)(12,0){5}{\circle*{2}}
\multiput(168,36)(12,0){5}{\circle*{2}}
\multiput(168,48)(12,0){5}{\circle*{2}}

\put(168,24){\line(1,-1){12}}\put(180,12){\line(1,1){12}}
\put(192,24){\line(1,-1){12}} \qbezier[8](204,12)(210,18)(216,24)

\put(168,48){\line(1,-1){12}}\put(180,36){\line(1,1){12}}
\put(192,48){\line(1,-1){12}} \qbezier[8](204,36)(210,30)(216,24)
\put(152,20){$L_1^{(3)}$} \put(154,46){$L_2^{(3)}$}
\put(224,27){$\rightarrow$}

\put(234,12){\line(1,0){48}}\put(246,0){\line(0,1){48}}
\multiput(246,12)(12,0){4}{\circle*{2}}
\multiput(246,24)(12,0){4}{\circle*{2}}
\multiput(246,36)(12,0){4}{\circle*{2}}
\multiput(246,48)(12,0){4}{\circle*{2}}

\put(246,24){\line(1,-1){12}} \put(258,12){\line(1,1){12}}
\put(270,24){\line(1,-1){12}} \put(246,48){\line(1,-1){12}}
\put(258,36){\line(1,1){12}} \put(270,48){\line(1,-1){12}}

\end{picture}
\end{center}
\caption{From 2-walks to 2-watermelons.\label{lower}}
\end{figure}
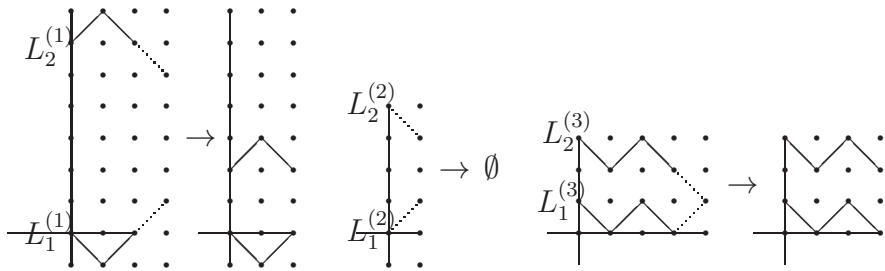

Observe that  by the choice of $j_k$, the rightmost pair of steps in
$(L_1^{(k)},L_2^{(k)})$ must be $(U,D)$. Moreover,  if we delete
this  pair of steps,  the resulting upper walk can be lowered
$2i-2k$ units without intersecting the lower walk to form a
$2$-watermelon. See Figure \ref{lower} for an example.

By Proposition \ref{watermelon}, The generating function for the
number of $2$-walks $(L_1^{(k)},L_2^{(k)})$ equals $D(t)=t\cdot
C^2(t)$. This completes the proof.\qed

Let $M(i,n)$ be the set of  intersecting 2-walks $(L_1,L_2)$ of
length $n$, where $L_1$ and $L_2$ start from $(0,0),(0,2i)$
respectively. Define
\begin{equation*}
 M_i(t)=\sum_{n\geq 0}|M(i,n)|t^n.
\end{equation*}

Observe that every pair of intersecting paths $(L_1, L_2)$ can be
decomposed into a pair of converging  paths and a pair of arbitrary
paths starting from the same point. Thus we have the following
formula.

\begin{cor}\label{intersect}
For any $i\geq 1$, \begin{equation*}M_i(t) ={D^i(t)\over
{1-4t}}.\end{equation*}
\end{cor}

It is obvious that
 \begin{equation}
 \label{twosum}M_i(t)+N_i(t)={1\over 1-4t}.\end{equation}
So the formula (\ref{tv2}) for $N_i(t)$ can be deduced from the
above formula. It is easy to see that $M_{12}(t)$, $M_{23}(t)$ and
$M_{13}(t)$ can be computed by using the above formula for $M_i(t)$.
So we get
\begin{equation}\label{gpq}
M_{12}(t)={D^i(2t)\over {1-8t}}, \quad M_{23}(t)={D^j(2t)\over
{1-8t}},\quad M_{13}(t)={D^{i+j}(2t)\over {1-8t}},
\end{equation}
in agreement with (\ref{w123}).
 Substituting (\ref{gpq}) into
(\ref{vm}), we obtain Theorem \ref{mainthm2}.

\section{Connection to the Ballot Numbers}

In this section, we put the Labelle merging algorithm in a more
general setting, and show that the direct correspondence formulated
by  Chen, Pang, Qu and Stanley \cite{CPQS} leads to a connection
between pairs of converging walks and the classical ballot numbers.

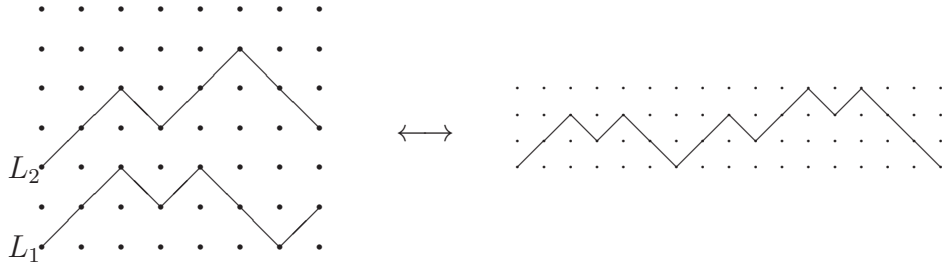
\begin{figure}[h,t]
\setlength{\unitlength}{1pt}  \begin{center}
\begin{picture}(330,100)  \multiput(0,0)(15,0){8}{\circle*{2}}
\multiput(0,15)(15,0){8}{\circle*{2}}
\multiput(0,30)(15,0){8}{\circle*{2}}
\multiput(0,45)(15,0){8}{\circle*{2}}
\multiput(0,60)(15,0){8}{\circle*{2}}
\multiput(0,75)(15,0){8}{\circle*{2}}
\multiput(0,90)(15,0){8}{\circle*{2}}
\put(0,0){\line(1,1){30}}\put(30,30){\line(1,-1){15}}
\put(45,15){\line(1,1){15}} \put(60,30){\line(1,-1){30}}
\put(90,0){\line(1,1){15}}

\put(-13,-4){$L_1$}\put(-13,26){$L_2$}

 \put(0,30){\line(1,1){30}}\put(30,60){\line(1,-1){15}}
\put(45,45){\line(1,1){30}} \put(75,75){\line(1,-1){30}}

 \put(134,40){$\longleftrightarrow$}
\multiput(180,60)(10,0){17}{\circle*{1}}
\multiput(180,50)(10,0){17}{\circle*{1}}
\multiput(180,40)(10,0){17}{\circle*{1}}
\multiput(180,30)(10,0){17}{\circle*{1}}


\put(180,30){\line(1,1){20}} \put(200,50){\line(1,-1){10}}
\put(210,40){\line(1,1){10}} \put(220,50){\line(1,-1){20}}
\put(240,30){\line(1,1){20}} \put(260,50){\line(1,-1){10}}
\put(270,40){\line(1,1){20}} \put(290,60){\line(1,-1){10}}
\put(300,50){\line(1,1){10}} \put(310,60){\line(1,-1){30}}
 \end{picture}  \end{center}
\caption{From a $2$-watermelon to a Dyck path.\label{water2D}}
\end{figure}

Let us recall the direct correspondence given in \cite{CPQS}. We
will represent a walk as a sequence of steps rather than points.
 Let
$(L_1, L_2)$ be a 2-watermelon of length $n$, and let
$L_1=p_1p_2\cdots p_n$ and $L_2= q_1q_2\cdots q_n$, where $p_i,
q_i=U$ or $D$. Set $U'=D$ and $D'=U$. Using the direct
correspondence in \cite{CPQS}, the watermelon $(L_1, L_2)$ can be
represented by a Dyck path of length $2n+2$:
\[ U q_1p_1' q_2 p_2'\cdots q_n p_n' D.\]
It is not difficult to see that the above correspondence is a
bijection. Figure \ref{water2D} gives an illustration.

Using the same idea, we may encode a pair of converging walks
$(L_1,L_2)$ in  $T(i,n)$ by a partial Dyck path $P$ in the sense
that the starting point of $P$ is not necessarily the point $(0,0)$.
We should note that the common definition of a partial Dyck path is
a lattice path starting from the origin $(0,0)$ with up and down
steps not going below the $x$-axis. Define $P(i,n)$ to be the set of
all partial Dyck paths of length $2n$ which start from $(0,2i)$ and
never return to the $x$-axis except for the final destination. The
following proposition establishes the connection between converging
walks and partial Dyck paths.

\begin{prop}
For $n\geq 1$, there exists a bijection between $T(i,n)$ and
$P(i,n)$.
\end{prop}

\begin{figure}[h,t]
\setlength{\unitlength}{1pt}
\begin{center}
\begin{picture}(340,130)
\put(-12,14){\line(1,0){108}} \put(0,2){\line(0,1){108}}

\multiput(0,12)(12,0){9}{\circle*{2}}
\multiput(0,24)(12,0){9}{\circle*{2}}
\multiput(0,36)(12,0){9}{\circle*{2}}
\multiput(0,48)(12,0){9}{\circle*{2}}
\multiput(0,60)(12,0){9}{\circle*{2}}
\multiput(0,72)(12,0){9}{\circle*{2}}
\multiput(0,84)(12,0){9}{\circle*{2}}
\multiput(0,96)(12,0){9}{\circle*{2}}
\multiput(0,108)(12,0){9}{\circle*{2}}
\multiput(0,120)(12,0){9}{\circle*{2}}

\put(0,24){\line(1,-1){12}}\put(12,12){\line(1,1){36}}
\put(48,48){\line(1,-1){12}}\put(60,36){\line(1,1){12}}
\put(72,48){\line(1,-1){12}}\put(84,36){\line(1,1){12}}

\put(0,96){\line(1,1){12}}\put(12,108){\line(1,-1){48}}
\put(60,60){\line(1,1){12}}\put(72,72){\line(1,-1){24}}
\put(-12,20){$L_1$}\put(-12,92){$L_2$}

\put(132,12){\line(1,0){204}} \put(144,0){\line(0,1){120}}
\put(133,81){$2i$}

\multiput(144,12)(12,0){17}{\circle*{2}}
\multiput(144,24)(12,0){17}{\circle*{2}}
\multiput(144,36)(12,0){17}{\circle*{2}}
\multiput(144,48)(12,0){17}{\circle*{2}}
\multiput(144,60)(12,0){17}{\circle*{2}}
\multiput(144,72)(12,0){17}{\circle*{2}}
\multiput(144,84)(12,0){17}{\circle*{2}}
\multiput(144,96)(12,0){17}{\circle*{2}}
\multiput(144,108)(12,0){17}{\circle*{2}}
\multiput(144,120)(12,0){17}{\circle*{2}}

\put(144,84){\line(1,1){24}}\put(168,108){\line(1,-1){84}}
\put(252,24){\line(1,1){24}}\put(276,48){\line(1,-1){24}}
\put(300,24){\line(1,1){12}}\put(312,36){\line(1,-1){24}}

\put(114,62){$\leftrightarrow$} \put(144,77){$P$}

\end{picture}
\caption{From a pair of converging walks to a partial Dyck
path.\label{merging}}
\end{center}
\end{figure}
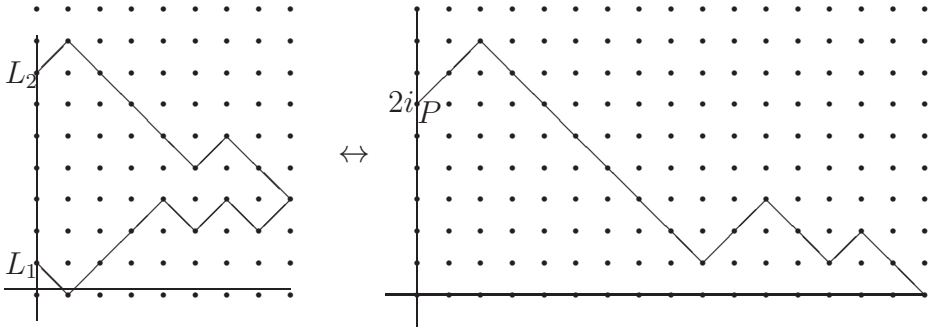

\pf Given a pair of converging walks $(L_1,L_2)$ in $T(i,n)$, let
$L_1=p_1p_2\cdots p_n$ and $L_2= q_1q_2\cdots q_n$, where $p_i,
q_i=U$ or $D$. Then $(L_1, L_2)$ can be represented by a partial
Dyck path $P$ of length $2n$ starting from $(0,2i)$:
\[P=q_1p_1' q_2 p_2'\cdots q_n p_n'.\]
Clearly, $P$ returns to the $x$-axis at the ending point and never
touches the $x$-axis before the ending point, that is, $P\in
P(i,n)$. It is easy to verify that the above correspondence is a
bijection. Figure \ref{merging} is an illustration.\qed

It is well known that the number of partial Dyck paths in $P(i,n)$
is given by the classical ballot number. Here we give a
decomposition
 of a partial Dyck path into Dyck paths in accordance with the
 generating function of $|T(i,n)|$ as given in Proposition
 \ref{b-prop}.

Given a partial Dyck path $P$ in $P(i,n)$, we can decompose $P$ into
$i$ nonempty Dyck paths $P_1,\ldots,P_{i}$ via the following
procedure. Let $P=A_0A_1\cdots A_{2n}$, where $P$ is represented by
the sequence of points rather than steps. Let $j_0=0$, and for
$1\leq k\leq i$, let $j_k$ be the minimum index such that the
$y$-coordinate of $A_{j_k}$ is two less than that of $A_{j_{k-1}}$.
Then we can decompose $P$ into $i$ segments $Q_1, Q_2, \ldots, Q_i$,
where $Q_k$ is the segment of $P$ starting at $A_{j_{k-1}}$ and
ending at $A_{j_k}$. Observe that by the choice of $j_k$, the
rightmost two steps of $Q_k$ must be $DD$. Let $P_k$ denote the Dyck
path obtained from $Q_k$ by deleting the last down step and adding
an up step before the first step of $Q_k$. Evidently, $P_k$ is a
nonempty Dyck path.  This completes the proof.\qed

To conclude this paper, we note that $|T(i,n)|$ can be computed by
using the Lagrange inversion formula, or by using the formula for
the number of Dyck paths of length $2n+2i$ with $2i$ returns to the
$x$-axis, see Deutsch  \cite{Deut}.  The explicit formula is as
follows:
\[|T(i,n)|={i\over n}{2n\choose n-i}.\]
We also note that  $|T(i,n)|$ can be expressed as the classical
ballot number
 $b(n+i-1,n-i)$, where
 \[ b(n,i)=
{n+i\choose i}-{n+i\choose i-1} ={n+1-i\over n+1+i}{n+i+1\choose
i},\]
see, for example, Riordan \cite{Rior}.

\vskip 3mm
 {\noindent \bf Acknowledgments.} We would like to thank  Ira Gessel
 for helpful discussions. This work was
supported by the 973 Project, the PCSIRT Project of the Ministry
of Education, the Ministry of Science and Technology, and the
National Science Foundation of China.

\end{document}